\documentclass[intlimits,twoside,a4paper]{article}
\usepackage{amsmath,amssymb}
\usepackage{graphicx}
\usepackage{wrapfig}
\usepackage[T2A]{fontenc}
\usepackage[cp1251]{inputenc}
\usepackage{cmpj}
\issue{2011}{14}{3}{33004}

\doinumber{10.5488/CMP.14.33004}

\def\be{\begin{equation}}
\def\ee{\end{equation}}
\def\bea{\begin{eqnarray}}
\def\eea{\end{eqnarray}}

\def\gsim{\mathrel{\lower.65ex\hbox{$\mathop{\kern0pt\sim}\limits
   ^{\lower.55ex\hbox{$>$}}$}}}
\def\lsim{\mathrel{\lower.65ex\hbox{$\mathop{\kern0pt\sim}\limits
   ^{\lower.55ex\hbox{$<$}}$}}}

\title[Improved MSA theory for simple fluids]
{Improved first order mean-spherical approximation \\ for simple fluids
}

\authorcopyright{S. Hlushak, A. Trokhymchuk, I. Nezbeda, 2011}
\author[S. Hlushak, A. Trokhymchuk, I. Nezbeda]
{S.~Hlushak\refaddr{ICMP},
A.~Trokhymchuk\refaddr{ICMP,BYU}, I.~Nezbeda\refaddr{PU}
}
\addresses{
\addr{ICMP} Institute for Condensed Matter Physics
of the National Academy of Sciences of Ukraine,\\
1 Svientsitskii Str., 79011 Lviv, Ukraine
\addr{BYU}
Department of Chemistry and Biochemistry, Brigham Young
University, Provo, UT 84602, USA
\addr{PU}
Faculty of Science,
J.E. Purkinje University, 400 96 Usti n. Lab., Czech Republic
}
\date{Received July 1, 2011, in final form July 25, 2011}

\begin{document}

\maketitle

\begin{abstract}

A perturbation approach based on the first-order mean-spherical
approximation (FMSA) is proposed. It consists in adopting a
hard-sphere plus short-range attractive Yukawa fluid as the novel
reference system, over which the perturbative solution of the
Ornstein-Zernike equation is performed. A choice of the optimal
range of the reference attraction is discussed. The results are
compared against conventional FMSA/HS theory and   Monte-Carlo
simulation data for compressibility factor and vapor-liquid phase
diagrams of the medium-ranged Yukawa fluid. The proposed theory keeps
the same level of simplicity and transparency as the conventional FMSA/HS
approach does, but turns out to be more accurate.
\keywords hard-core Yukawa fluid, mean-spherical approximation,
perturbation theory, Monte Carlo simulations
\pacs 01.65.Q
\end{abstract}

\section{Introduction}

The most common and universal method in theories of fluids is the perturbation approach. Nonetheless, without analytic results of the integral equation theory for simple systems the perturbation expansion could not be accomplished. To be specific, the analytic results of the mean-spherical approximation (MSA) for the fluid of hard spheres (for which the MSA is identical to the Percus-Yevick theory)~\cite{WertheimHS,BaxterHS} and the Yukawa fluid~\cite{analMSA,blumhoye} are used to describe the properties of the reference system. In addition to these conventional MSA results there is also a perturbed version of the MSA
called the first-order mean-spherical approximation (FMSA)~\cite{tang1}.

When compared to computer simulation data for the Yukawa fluid
model, the FMSA theory has been shown to be reasonably good but
still less accurate than the common MSA~\cite{tang2000}. However, an advantage of the FMSA theory, hereinafter referred to as FMSA/HS since it employs the hard sphere (HS) fluid for the reference, is its simplicity and transparency. These are the factors causing its increasing attractiveness and potential applicability in the liquid state theory and also, in general, in the soft condensed matter theory.
The most important ideas of the FMSA/HS theory are outlined in the following section.

The goal of this paper is to present a modification of the FMSA/HS theory. The modification pursues the idea that there is an alternative to the HS reference, namely a mean field theory based on a Yukawa fluid reference~\cite{melnyk4}.
In section~3 the possible choices in this direction are considered.
The results and discussions are the subject of section~4. The study is summarized in section~5.

\section{Conventional FMSA/HS theory}

The FMSA/HS is a theory based on the solution of the
Ornstein-Zernike (OZ) equation,
\begin{equation}
\tilde{h}\left( k\right) =\tilde{c}\left( k\right) +\rho \tilde{h}\left(
k\right) \tilde{c}\left( k\right)
\label{oz}
\end{equation}
by employing the perturbation expansions limited to the first
order correction term,
\begin{eqnarray}
\tilde{h}\left( k\right) &=&\tilde{h}_{\rm o}\left( k\right) + \Delta \tilde{h}\left( k\right)
,\nonumber \\
%+\dots  \label{h_pert} \\
\tilde{c}\left( k\right) &=&\tilde{c}_{\rm o}\left( k\right) + \Delta \tilde{c}\left( k\right)
%+\dots ,
,\nonumber\\
\hat{Q}\left( k\right) &=&\hat{Q}_{\rm o}\left( k\right) + \Delta \hat{Q}\left( k\right)\,,
%+\dots ,
%\label{Q_pert}
\label{c_pert}
\end{eqnarray}
for the total, $h$, and direct, $c$, correlation functions
combined with Baxter factorization function, $Q$,
respectively. Here and in what follows all symbols with the tilde
will denote the three-dimensional Fourier transforms, while all
symbols with the hats will denote the one-dimensional Fourier
transforms or the Laplace transforms. The terms with subscript
``${\rm o}$'' denote the contribution of a reference system, and
$\Delta h$, $\Delta c$, $\Delta Q$ are the respective
corrections.

The FMSA/HS theory has been developed by Tang and Lu
 \cite{tang1,tang2,tang3} for the hard-core based fluid models
 \begin{equation}
u\left( r\right) =\left\{
\begin{array}{ll}
\infty , & \quad r < \sigma, \\
\phi (r), & \quad r \geqslant  \sigma,
\end{array}
\right.  \label{Yukpot}
\end{equation}
with $\sigma$ being the diameter of a hard core and
$\phi(r)$ being the potential function for out-of-core pair
interaction between molecules. In their studies Tang and Lu were
considering the fluid of hard spheres (HS) as the reference (or
unperturbed) system for which all three functions that enter
expansions (\ref{c_pert})
are well known. In particular, for the total correlation function
of the HS reference system, $h_{\rm o}\equiv h_{\rm HS}$\,,
Tang and Lu used the expression
\begin{equation}
  \hat{g}_{\rm o}\left( s \right) \equiv \hat{g}_{\rm HS}\left( s \right) =
  \frac{L\left( s\sigma \right)\re^{-s\sigma}}{\left( 1-\eta \right)^2
  \hat{Q}_{\rm HS}\left( s\sigma \right)s^2} \,,
  \label{g10}
\end{equation}
that is the Laplace transform
%$\hat{g}_{\rm HS}(s)$
of the radial distribution function $g_{\rm HS}(r)=h_{\rm HS}(r)+1$ resulting from the  solution
of the OZ equation for HS fluid in the Percus-Yevick approximation \cite{WertheimHS}.
The function $\hat{Q}_{\rm HS}\left( t\right)$ refers to the Laplace transform of the Baxter
factorization function of the HS fluid,
\begin{equation}
\hat{Q}_{\rm o}\left( t\right) \equiv \hat{Q}_{\rm HS}\left( t\right) =\frac{S\left( t\right) +12\eta
L\left( t\right) \re^{-t}}{\left( 1-\eta \right) ^{2}t^{3}}\,,
\label{Q_0}
\end{equation}
where
\begin{eqnarray}
S\left( t\right) &=&\left( 1-\eta \right) ^{2}t^{3}+6\eta \left(
1-\eta
\right) t^{2}+18\eta ^{2}t-12\eta \left( 1+2\eta \right) ,  \label{S_def} \\
L\left( t\right) &=&\left( 1+\frac{\eta }{2}\right) t+1+2\eta
\label{L_def}
\end{eqnarray}
and $\eta = \pi\rho\sigma^3/6$ is the packing fraction.
% parameter for a fluid of hard spheres.

The  main result obtained by Tang and Lu concerns an expression
for correction $\Delta h$ to the total correlation function
(or radial distribution function) resulting from the out-of-core
attraction $\phi(r)$. It reads~\cite{tang1}
\begin{equation}
%\varepsilon
\Delta \hat{h}\left( k\right) =\frac{P\left( \ri k\sigma\right)
}{{\hat{Q}_{\rm HS}^{2}\left( \ri k\sigma\right) }}\,,  \label{P_def}
\end{equation}
where
\begin{equation}
P\left( \ri k\sigma\right) =\frac{\Delta U\left( k\right)
}{2\hat{Q}_{\rm HS}^{2}\left(
-\ri k\sigma\right) }-\frac{\re^{-\ri k\sigma}}{2\ri\pi }\int_{-\infty }^{\infty }\frac{%
\Delta U\left( y\right) \re^{\ri y\sigma}}{\left( y-k\right)
\hat{Q}_{\rm HS}^{2}\left( -\ri y\sigma\right) }\rd y\,,  \label{P_k}
\end{equation}
with function $\Delta U\left( k\right)$ defined as
\begin{equation}
\Delta U\left( k\right) = \int_{\sigma}^{\infty } r\Delta c\left( r\right) \re^{-\ri kr}\rd r.
\label{U_1_def}
\end{equation}
In accordance with this approach the necessary closure to be used
in equation (\ref{U_1_def}) reads as
\begin{equation}
\Delta c^{\rm FMSA/HS}\left( r\right) =-\beta \phi\left( r\right) ,\qquad \mathrm{for}\qquad r\geqslant \sigma,
\label{c_1_defHS}
\end{equation}
with $\beta =1/k_{\mathrm{B}}T$ and $T$ being the temperature.

In general, without any relation to the above version of the FMSA theory, a key point of
its application to particular fluid model [determined by the out-of-core interaction
potential $\phi(r)$] concerns an evaluation of the integral
in equation~(\ref{P_k}). So far, within the framework of the FMSA/HS theory
this has been done for the Yukawa, Lennard-Jones, Kihara and sticky
fluids  by Tang and Lu~\cite{tang2} and for the square-well fluids by Tang and Lu~\cite{tang3}
and by Hlushak et al.~\cite{hlushak1}.
In the particular case of the Yukawa (Y) fluid model,
\begin{equation}
\phi_{\rm Y}\left( r\right) =
-\epsilon\sigma\frac{\re^{-z(r-\sigma)}}{r}\,, \label{uyuk}
\end{equation}
for which the function $\Delta U\left(k\right) \sim \re^{-\ri k\sigma}$, an
integration contour in the right-hand side of equation~(\ref{P_k}) can
be closed in the upper complex half-plane and evaluation of the
integral is rather simple and does not require the calculation of the residues at zeroes of
the function $\hat{Q}_{\rm HS}\left( -\ri k\sigma\right)$. Then,
the Laplace transform of the correction term
(\ref{P_def}) for the total correlation function of the Yukawa fluid
 within the FMSA/HS theory reads
\begin{equation}
  \Delta \hat{h}^{\rm FMSA/HS}\left(s\right) =
  \frac{\beta\epsilon\sigma \re^{-s\sigma}}{\left(s+z\right)\hat{Q}^2_{\rm HS}
  \left(s\sigma \right)\hat{Q}^2_{\rm HS}\left(z\sigma \right)}\,.
  \label{h1s}
\end{equation}

Once the radial distribution function $\hat{g}=\hat{g}_{\rm
o}+\Delta \hat{h}$ is known, the thermodynamics of the system
can be calculated.

\section{FMSA theory based on Yukawa reference system}

The first order mean-spherical approximation theory that we are dealing with is a kind of
 perturbation theory approach. Usually,
within the perturbation theory, in order to improve the performance of the first
order approximation one should calculate the second order correction term.
As an alternative to this common way
here we propose to qualitatively modify  the reference system over which the perturbation is
calculated, and as a result to continue working within the same first order approximation.
This is of particular importance since our aim is to keep the simplicity and transparency
of the improved FMSA theory at the level established by the conventional FMSA/HS approach.

\subsection{Division of the potential}
Similar to the classical perturbation theory approach we proceed
by dividing the initial interaction potential $u(r)$ into two parts
\begin{equation}
  u\left( r \right) =   u_{\rm o}\left( r \right) + \Delta u\left( r \right),
  \label{urdiv}
\end{equation}
that are the reference and residual contributions, respectively.
However, for the reference system potential $u_{\rm o}\left( r \right)$ we will require
that besides the hard-core repulsion it should
(i) include a piece of the attractive tail of the same strength $\epsilon$ as the total
potential, and
(ii) extend over the range that is somewhat shorter than the
range of the total interaction. The function that allows us to satisfy and control these
requirements in an easy and a natural way is the Yukawa (Y0) potential.
Thus, we may define the desired reference fluid as follows:
 \begin{equation}
   u_{\rm o}\left( r\right) \equiv u_{\rm Y0}(r) =  \left\{
\begin{array}{ll}
\infty , & \quad r < \sigma ,\\
-\varepsilon\sigma {\re^{-z_{\rm o}\left( r-\sigma \right)}}/{r}, & \quad r \geqslant  \sigma\,.
\end{array}
\right.  \label{Yukpotprime}
\end{equation}

The improved FMSA closure, referred to as FMSA/Y0, reads as follows:
\begin{equation}
\Delta c^{\rm FMSA/Y0}\left( r\right) =-\beta \phi\left( r\right)-
\beta\epsilon\sigma {\re^{-z_{\rm o}\left( r-\sigma
\right)}}/{r},\qquad \mathrm{for}\qquad r\geqslant \sigma\,,
\label{c_1_defY0}
\end{equation}
where $z_{\rm o}$ is the so far undefined parameter.

\subsection{Yukawa reference system}
We note here that some authors have already suggested to utilize the non-HS reference system
in liquid state theory (e.g., see~\cite{ref1,ref2}).
Moreover, the short-range attractive Yukawa fluid has been already
considered as an alternative
to the HS reference system by Melnyk et al.~\cite{melnyk4,melnyk1,melnyk2,melnyk3,melnyk5}
in their studies within the framework of an augmented van der Waals theory for simple fluids.

It is very important that, like in the case of the HS reference potential, all properties
for the Yukawa reference potential (\ref{Yukpotprime}), including the Baxter function
$Q_{\rm o}\equiv Q_{\rm Y0}$\,, are available in the literature.
%The properties of the Y0 reference system, including Baxter function $Q_0$
They can be obtained either within the MSA theory for Yukawa fluid
[e.g., see Blum and Hoye~\cite{blumhoye}, Kalyuzhnyi et al.~\cite{yukalyuk,yukalyuk2}]
or within the conventional FMSA/HS theory due to Tang and Lu as it is described in previous
section~2.
The MSA theory is more accurate than FMSA/HS theory, but the latter is simpler.
For the purpose of the present study we decided to sacrifice the accuracy in order to maintain
simplicity and transparency.
Thus, in what follows the FMSA/HS theory will be employed to describe properties of the
Y0 reference system.
And, as we will see  hereinafter, despite a less accurate description of the reference system,
the FMSA/Y0
theory still shows a notable improvement against the FMSA/HS theory.

Following the conventional  FMSA/HS approach,  the Laplace transform of the radial
distribution function of
the Y0 reference system reads
\begin{equation}
  \hat{g}_{\rm o}\left( s \right) \equiv \hat{g}_{\rm Y0}\left( s;z_{\rm o} \right) =
  \hat{g}_{\rm HS}\left( s \right) +
\frac{\beta\epsilon\sigma \re^{-s\sigma}}{\left( s+z_{\rm o} \right)
\hat{Q}^2_{\rm HS}\left( s\sigma \right)\hat{Q}^2_{\rm HS}\left( z_{\rm o}\sigma \right)}\,,
  \label{gY010}
\end{equation}
where the first term corresponds to the contribution of
a hard-sphere repulsion, while the second one is the contribution
due to the short-range attraction attributed to the Y0 reference
system in accordance with equation~({\ref{Yukpotprime}). Similarly, the FMSA/HS result for
the Baxter factorization function of the Y0 reference is
% \cite{tang2003first}
%
\begin{equation}
  \hat{Q}_{\rm o}\left( t \right) \equiv
  \hat{Q}_{\rm Y0}\left( t;z_{\rm o}\sigma \right) = \hat{Q}_{\rm HS}\left(t \right) +
  12\beta\varepsilon\eta\frac{z_{\rm o}\sigma \hat{Q}_{\rm HS}\left( -t \right)\re^{-t}
  -\left( t+z_{\rm o}\sigma \right)\hat{Q}_{\rm HS}\left( t \right)}
  {tz_{\rm o}\sigma\left( t+z_{\rm o}\sigma \right)\hat{Q}^2_{\rm HS}\left( z_{\rm o}\sigma \right)}\,.
  \label{Q10}
\end{equation}

\subsection{Full system within the FMSA/Y0 theory}
After all properties of interest for the Y0 reference fluid are
specified, we turn to the entire system determined by the
interaction potential $u(r)$ or more precisely by the
out-of-core potential function $\phi(r)$. Although this
function can be substituted by any potential function used in
literature to represent the simple fluids [e.g., Lennard-Jones,
Sutherland, Yukawa or Kihara potentials, etc.] for the purpose of
present study we proceed with the Yukawa (Y) fluid model already
defined according to equation~(\ref{uyuk}).

By introducing  Yukawa potential $\phi_{\rm Y}(r)$ into the
FMSA/Y0 closure (\ref{c_1_defY0}) we note that for both functions
that enter $\Delta c^{\rm FMSA/Y0}(r)$, the function $\Delta
U\left(k\right)$ is proportional to $\re^{-\ri k\sigma}$ and the
integral in equation~(\ref{P_k}) can be easily evaluated in the way it was already discussed in Introduction section. Then, the
correction term $\Delta \hat{h}^{\rm FMSA/Y0}$ that is
necessary to evaluate the radial distribution function
%$\hat{g}=\hat{g}_{\rm Y0}+\Delta \hat{h}$
of the full system,
\begin{equation}
  \hat{g}\left( s;z_{\rm o},z \right)=\hat{g}_{\rm o}\left( s;z_{\rm o} \right) + \Delta\hat{h}\left( s;z_{\rm o},z \right),
  \label{hfull}
\end{equation}
%
%for the Yukawa fluid of interest
is given by
\begin{eqnarray}
  \Delta\hat{h}^{\rm FMSA/Y0}\left( s;z_{\rm o},z \right) &=&
  -\frac{\beta\epsilon\sigma \re^{-s\sigma}}{\left( s+z_{\rm o} \right)\hat{Q}^2_{\rm Y0}\left( s\sigma;z_{\rm o}\sigma
\right)\hat{Q}^2_{\rm Y0}\left( z_{\rm o}\sigma;z_{\rm o}\sigma \right)} \nonumber\\
 &&{} +\frac{\beta\epsilon \sigma \re^{-s\sigma}}{\left( s+z \right)\hat{Q}^2_{\rm Y0}\left( s\sigma;z_{\rm o}\sigma \right)
\hat{Q}^2_{\rm Y0}\left( z\sigma;z_{\rm o}\sigma \right)}\,.
  \label{h11}
\end{eqnarray}

The calculations of thermodynamics within the FMSA/Y0 approach
can be made through the energy route, i.e., in the way that is
quite similar to how it was done by Tang et al. within the
conventional FMSA/HS theory \cite{tang2000}. It consists in
evaluating the internal energy using its definition through the
radial distribution function,
\begin{equation}
  \frac{ U}{NkT} = 2\pi\rho\beta \int_0^{\infty}\rd r r^2
  g\left( r;z_{\rm o},z \right)u\left( r \right)
  = 12\eta\beta\varepsilon\sigma  \re^{z\sigma} \hat{g}\left( z;z_{\rm o},z
  \right).
  \label{UnaN}
\end{equation}
This result is used to derive the Helmholtz free energy in
the form
\begin{equation}
  \frac{A-A_{\rm id}}{NkT}=a_0 + a_1 + a_2\,,
  \label{Afull}
\end{equation}
where
\begin{eqnarray}
  a_0 &=& \frac{4\eta-3\eta^2}{\left( 1-\eta
  \right)^2}\,,\label{a0}\\
  a_1 &=&  -12\eta\frac{\beta\varepsilon L\left( z\sigma \right)}
  {\left( 1-\eta \right)^2\hat{Q}_{\rm HS}\left( z\sigma \right){\left( z\sigma \right)}^2}\,, \label{a1}\\
  a_2 &=& -6\eta\beta^2\varepsilon^2\left[\frac{1}
  {\left( z\sigma+z_{\rm o}\sigma \right)\hat{Q}^2_{\rm HS}\left( z\sigma \right)\hat{Q}^2_{\rm HS}\left( z_{\rm o}\sigma \right)}
  \right.\nonumber\\
  &&{}-\frac{1}{\left( z\sigma+z_{\rm o}\sigma \right)\hat{Q}^2_{\rm Y0}
  \left( z\sigma;z_{\rm o}\sigma \right)\hat{Q}^2_{\rm Y0}\left( z_{\rm o}\sigma;z_{\rm o}\sigma \right)}+\left.\frac{1}{ 2z\sigma\hat{Q}^2_{\rm Y0}\left( z\sigma;z_{\rm o}\sigma \right)
  \hat{Q}^2_{\rm Y0}\left( z\sigma;z_{\rm o}\sigma \right)}
  \right]\,.\label{a2}
\end{eqnarray}
In the limit $z_{\rm o}=z$ the above formulas reduce to the
conventional FMSA/HS theory by Tang and Lu \cite{tang2000}. The
chemical potential and pressure are obtained employing standard
thermodynamic relations.

\section{Results and discussions}

\subsection{How much of the attraction should be treated as the reference?}

\begin{wrapfigure}{o}{0.5\textwidth}
\vspace{-0.4cm}
    \includegraphics[width=7cm]{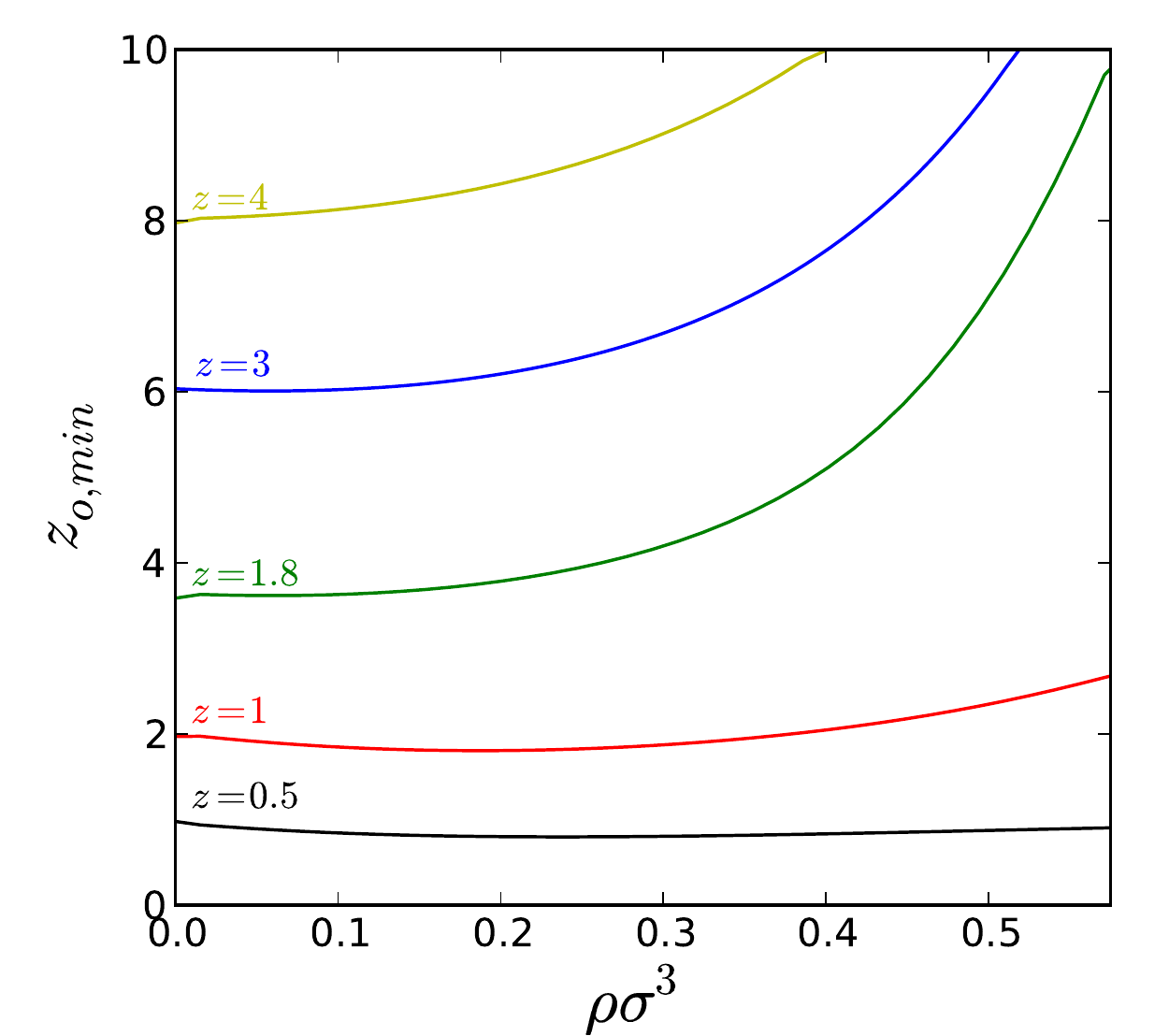}
  \caption{ Density dependence of the Y0 reference decay parameter $z_{\rm o,min}$
            that minimizes the free energy for Yukawa fluid models
            with different value of the decay exponent $z$ as it is specified for each
            curve. One can see that at low density $z_{\rm o,min}\to 2z$. }
  \label{fig:zmin}
  \vspace{-1ex}
\end{wrapfigure}
The above outlined FMSA/Y0 theory contains the parameter $z_{\rm
o}$ that determines the range of reference attraction and needs
to be specified. The most straightforward approach to proceed
deals with the minimization of the free energy~(\ref{Afull}) of the
Yukawa system over the range of possible values of $z_{\rm
o}$\,. Noting that $z_{\rm o}$ enters only $a_2$ contribution,
we deduce that the outcome of the minimization, i.e., $z_{\rm o,
min}$\,, depends only on the density of the fluid but not on the
temperature. The corresponding density dependence of $z_{\rm o,
min}$ for several Yukawa fluids is presented in
figure~\ref{fig:zmin}. It is evident that in the
limit of a vanishing density and for all systems considered, the
parameter $z_{\rm o, min}$ tends to the value of $2z$.

This rather remarkable result for choosing the Y0 reference system in such a strict form is quite probably
limited to the FMSA theory only. Nevertheless, it is quite consistent with our earlier findings for the
Y0 reference decay parameter within the framework of the Yukawa
based van der Waals theory for the simple fluids~\cite{melnyk4,melnyk1,melnyk2,melnyk3,melnyk5}.
First of all, there is a requirement for $z_{\rm o,min}$ to be larger than $z$ providing in this way
the range of the reference attraction which will be shorter than that in the parent fluid.
Secondly, it is obvious that the reference fluid should be in a one-phase region for the set of density
and temperature parameters used in the studies of the parent fluid. In the case of a hard sphere reference,
this requirement is satisfied since HS fluid does not exhibit the liquid/vapor transition. By adding an attraction one
 provides the possibility for the phase coexistence. However, the critical temperature is always getting
lower if the range of attraction is shorter. In~\cite{melnyk2} we presented a collection of the
Monte Carlo generated liquid/vapor envelopes for Yukawa fluid with different values
of the decay parameter (see figure~1 in~\cite{melnyk2}) from which it follows that the values
$3<z_{\rm o}\sigma<6$ can be used as a reference decay for the Lennard-Jones-like Yukawa fluid
($z\sigma=1.8$) since the critical point temperature in such a reference fluid will always
 be lower than the triple point temperature in parent fluid.

On the other hand, analyzing the dependence of the critical point coordinates of the Lennard-Jones-like Yukawa fluid ($z\sigma=1.8$) on the
reference system decay parameter $z_{\rm o}$\,, in~\cite{melnyk1} we have shown that the
best agreement with computer simulation data can be reached if $3<z_{\rm o}\sigma<4$.
Thus, the result $z_{\rm o}=2z$ agrees with these findings and in all our
subsequent calculations for the Lennard-Jones-like Yukawa fluid we impose $z_{\rm o}\sigma=3.6$, making it
independent of the fluid density.
%We will leave investigation of the more general case of the density dependent $z_{\rm o, min}$ for future.

\subsection{Comparison between FMSA/HS and FMSA/Y0 solutions}
To illustrate the improvements that the FMSA/Y0 theory yields
over the conventional \linebreak FMSA/HS theory, we compare the predictions
made up by these two theoretical approaches
for compressibility factor and vapor/liquid phase diagram of the most popular
Yukawa fluid model defined by the decay parameter
$z\sigma=1.8$.

\begin{figure}[ht]
\includegraphics[width=7cm]{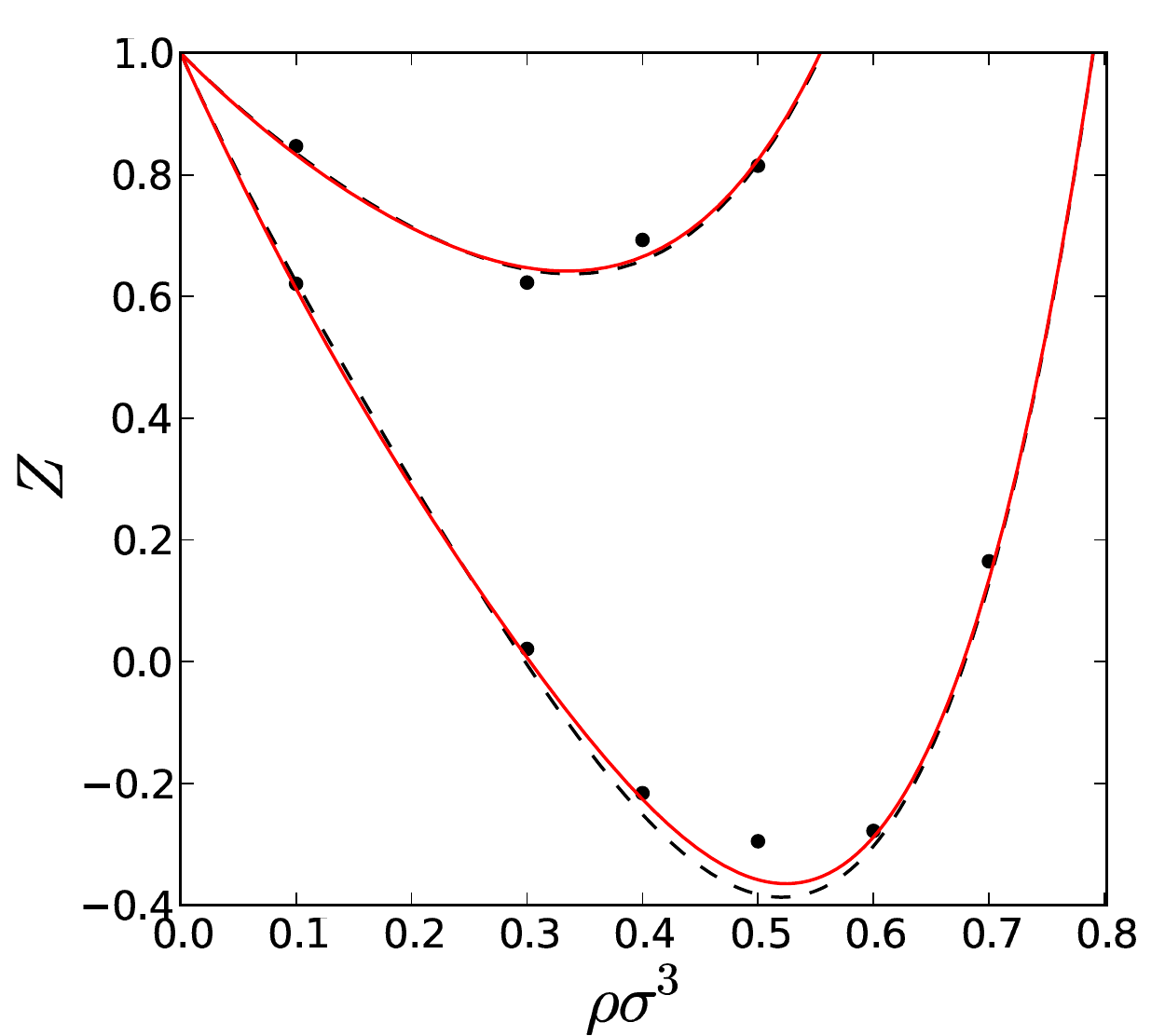}%
\hfill%
\includegraphics[width=7cm]{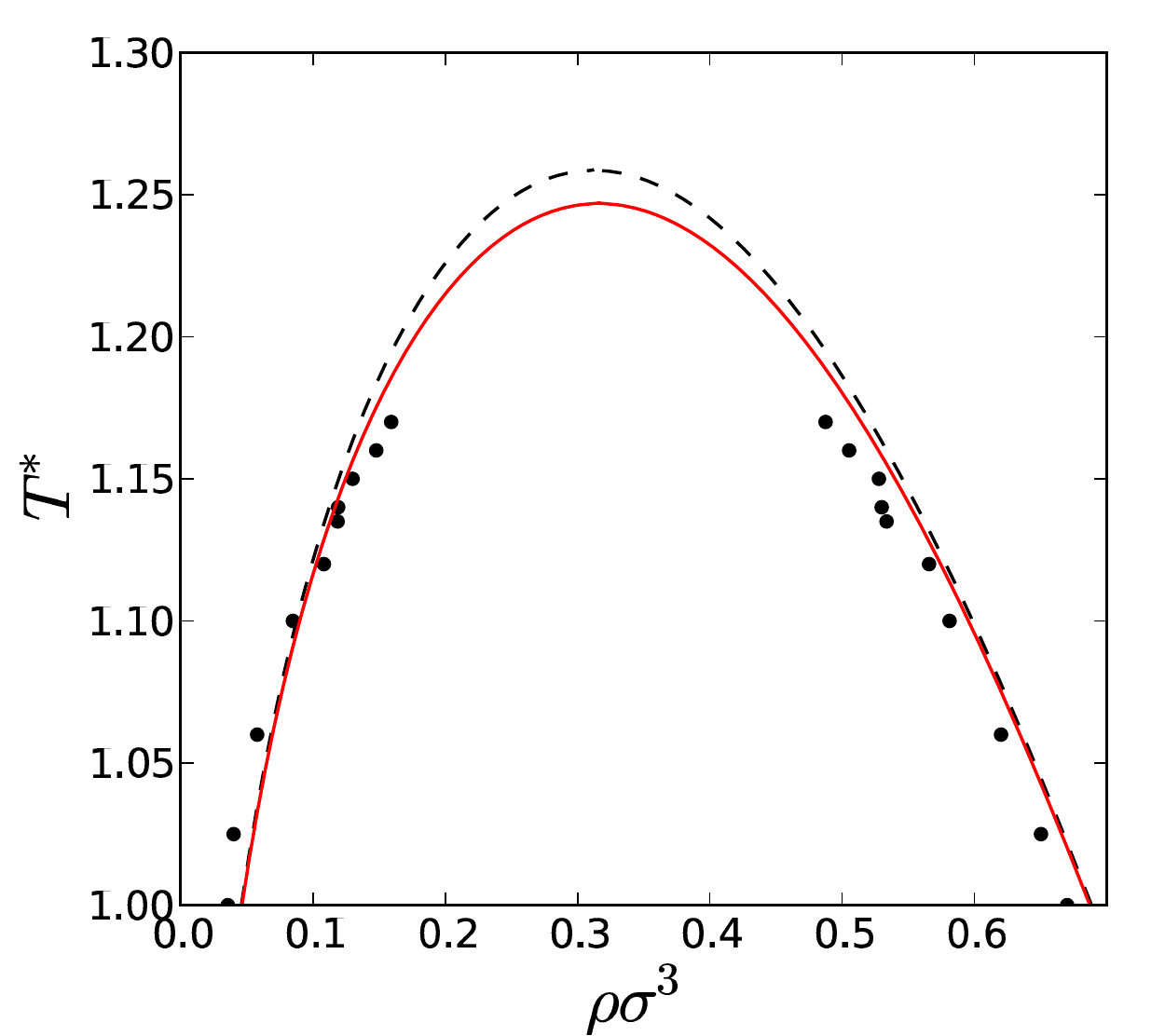}%
\\%
\parbox[t]{0.4808\textwidth}{%
\caption{Isotherms $T^*=1.5$ and $T^*=1$ (from the top to the bottom)
for compressibility factor, $Z=PV/NkT$, of the Yukawa fluid with
  $z\sigma=1.8$ as obtained from the conventional FMSA/HS theory (black dashed lines) and
  from the improved FMSA/Y0 theory with $z_{\rm o}\sigma=3.6$ (red solid lines).
  The symbols denote computer simulation data by Schukla~\cite{shukla2000phase}.
  }
  \label{fig:zz18}%
}%
\hfill%
\parbox[t]{0.48\textwidth}{%
\caption{Vapor-liquid phase diagram of the Yukawa fluid with
  $z\sigma=1.8$. The notations are the same as in~figure~\ref{fig:zz18}.}
  \label{fig:pdz18}%
}%
\end{figure}
The compressibility factors, $Z=PV/NkT$, of the Yukawa fluid with
$z\sigma=1.8$ and for two temperatures, $T^*=1.5$ and 1, that result from both
FMSA approaches are presented in figure~\ref{fig:zz18}. To estimate the effectiveness
of the theory, the computer simulation data due to Schukla~\cite{shukla2000phase}
are shown as well.
As it was already pointed out by Tang~\cite{tang2000},
the conventional FMSA/HS theory does a good job in predicting
the compressibility factor of the Yukawa fluid. Nevertheless, the
proposed FMSA/Y0 theory shows to be even more accurate, especially
at lower temperature and in the region of intermediate densities.

While the enhancement may seem to be minor in the case of compressibility
factor, it is more pronounced for the vapor/luqiud phase
diagram shown in figure~\ref{fig:pdz18}.
The results of the FMSA/Y0 theory lead to the shrinkage of coexisting
densities envelope in the region near the critical point, approaching the computer
simulation data. The estimated improvement in comparison with the FMSA/HS predictions
is around 30\%.

\section{Conclusions}

Being compared with  computer simulation data for the
Lennard-Jones-like Yukawa fluid ($z\sigma=1.8$), the first
order mean-spherical approximation (FMSA/HS) due to Tang et Lu has
shown to be only slightly less accurate in the calculations of the
thermodynamics and liquid/vapor phase coexistence than the full
MSA theory (details of this comparison and a corresponding
discussion can be found in the~\cite{tang2000}). At the same
time, the FMSA/HS is a much  simpler theory. Abbreviation HS
underlines here that the theory is based on the hard-sphere reference
system. In the present study we reported a further improvement of
this approach, which we called a conventional FMSA/HS theory by introducing
the short-ranged Yukawa fluid (with decay parameter $z_{\rm
o}=2z$) as a new reference system. Consequently, we are
referring to this theory as the FMSA/Y0 theory. Numerical
calculations that we performed for the same model discussed by Tang~\cite{tang2000} showed that the proposed
modifications of the reference system make the first order mean-spherical approximation even more accurate.

It is important that in order to treat the novel reference we are employing
the conventional FMSA/HS approach. Due to this, the level of
simplicity and transparency of the improved \linebreak FMSA/Y0 theory is kept
at the same level as that of the conventional FMSA/HS theory. In
particular, it is easy to see that the resulting new expressions for
main ingredients of the FMSA ideology~-- the Laplace transforms of
the radial distribution function, equation~(\ref{gY010}), and the
Baxter factorization function, equation~(\ref{Q10}), of the new
reference system, as well as the equation (\ref{h11}) for the
correction term -- are only slightly longer than their original
counterparts [see equations~(\ref{g10}),~(\ref{Q_0}) and (\ref{h1s})]
and both being composed of the same variables, functions and model
parameters. Only the parameter of the FMSA/Y0 theory appears to be
the decay parameter $z_{\rm o}$, for which in the case of
Yukawa fluid we obtained $z_{\rm o}=2z$,  and which determines
an amount of the out-of-core attraction that should be
attributed to the Y0 reference system.

\section*{Acknowledgement}

This work was supported by the Grant Agency of the Academy of Sciences
of the Czech Republic (Grant No. IAA400720710) and the Czech-Ukrainian Bilateral Cooperative Program.

%\newpage

\ukrainianpart

\title{Удосконалене середньосферичне наближення першого порядку для простих рідин}

\author{С.~Глушак\refaddr{ICMP},
А.~Трохимчук\refaddr{ICMP,BYU}, І.~Незбеда\refaddr{PU}
}
\addresses{
\addr{ICMP} Інститут фізики конденсованих систем НАН України,  вул. І. Свєнціцького, 1, 79011 Львів, Україна
\addr{BYU}
Факультет хімії та біохімії, Університет Бригам Янг, Прово, США
\addr{PU}
Факультет природничих наук, Унiверситет Я.Е. Пуркiнйє, Чеська Республiка
}

\makeukrtitle

\begin{abstract}
\tolerance=3000
Пропонується новий підхід теорії збурень на основі середньосферичного наближення першого порядку (ССНПП).
Він полягає у використанні твердосферної (ТС) рідини Юкави в якості базисної системи,
на основі якої розв'язується рівняння Орнштейна-Церніке
в рамках теорії збурень. У роботі обговорюється вибір оптимального параметра
притягальної далекодії базисної системи.
Результати порівнюються зі звичайною теорією ССНПП/ТС та моделюванням Монте-Карло,
для коефіцієнта стисливості і фазової діаграми рідини Юкави.
Запропонована теорія зберігає
попередній рівень простоти та прозорості як і звичайне ССНПП/ТС,
але є більш точною.

\keywords рідина Юкави, середньосферичне наближення,
теорія збурень, моделювання Монте-Карло
\end{abstract}

\end{document}